\documentclass[aps,pre,reprint,amsmath,amssymb,superscriptaddress,showkeys]{revtex4-1}

\usepackage[dvipsnames]{xcolor}
\usepackage[utf8]{inputenc}
\usepackage{amsmath, amsfonts, amsthm, bbm}
\usepackage{amsthm}
\usepackage{mathtools}
\usepackage{amssymb}
\usepackage{pifont}
\usepackage{bm}
\usepackage{graphicx} 
\usepackage{bmpsize}
\usepackage{tikz}
\usepackage{float}
\usepackage{braket}
\usepackage{hyperref}

\definecolor{brickred}{rgb}{0.8, 0.25, 0.33}
\newcommand\myshade{85}
\colorlet{mylinkcolor}{BrickRed}
\colorlet{mycitecolor}{NavyBlue}
\colorlet{myurlcolor}{Aquamarine}

\hypersetup{
  linkcolor  = mylinkcolor!\myshade!black,
  citecolor  = mycitecolor!\myshade!black,
  urlcolor   = myurlcolor!\myshade!black,
  colorlinks = true,
}

\newcommand{\tens}[1]{%
  \mathbin{\mathop{\otimes}\limits_{#1}}%
}

\def\oinfo{O-information}
\def\qinfo{Q-information}

\begin{document}

\title{Quantifying High-Order Interdependencies in Entangled Quantum States}

\author{Marco Alberto Javarone}
\email{marcojavarone@gmail.com}
\affiliation{Dipartimento Interateneo di Fisica, Università degli Studi di Bari Aldo
Moro Bari, I-70126 Bari, Italy}
\affiliation{Dutch Institute for Emergent Phenomena, Amsterdam, Netherlands}

\author{Fernando E. Rosas}
\affiliation{Department of Informatics, University of Sussex, Brighton, UK}
\affiliation{Centre for complexity science, Imperial College London, London, UK}
\affiliation{Centre for psychedelic research, Department of Brain Sciences, Imperial College London, London, UK}
\affiliation{Centre for Eudaimonia and human flourishing, University of Oxford, Oxford, UK}

\author{Paolo~Facchi}
\affiliation{Dipartimento Interateneo di Fisica, Università degli Studi di Bari Aldo
Moro Bari, I-70126 Bari, Italy}
\affiliation{INFN, Sezione di Bari, I-70126 Bari, Italy}

\author{Saverio Pascazio}
\affiliation{Dipartimento Interateneo di Fisica, Università degli Studi di Bari Aldo
Moro Bari, I-70126 Bari, Italy}
\affiliation{INFN, Sezione di Bari, I-70126 Bari, Italy}

\author{Sebastiano Stramaglia}
\email{sebastiano.stramaglia@ba.infn.it}
\affiliation{Dipartimento Interateneo di Fisica, Università degli Studi di Bari Aldo
Moro Bari, I-70126 Bari, Italy}
\affiliation{INFN, Sezione di Bari, I-70126 Bari, Italy}
\begin{abstract}
\noindent
Here, we leverage recent advances in information theory to develop a novel method to characterise the dominant character of the high-order dependencies of quantum systems. 
To this end, we introduce the \emph{Q-information}: an information-theoretic measure capable of distinguishing quantum states dominated by synergy or redundancy. 
We illustrate the measure by investigating the properties of paradigmatic entangled Qubit states and find that --- in contrast to classical systems --- quantum systems need at least four variables to exhibit high-order properties. 
Furthermore, our results reveal that unitary evolution can radically affect the internal information organisation in a way that strongly depends on the corresponding Hamiltonian.
Overall, the \qinfo~sheds light on novel aspects of the internal organisation of quantum systems and their time evolution, opening new avenues for studying several quantum phenomena and related technologies.
\end{abstract}
\maketitle

Entropy connects the physics of a system with the information it carries, being a fundamental quantity for describing the relationships among the degrees of freedom in a system. 
However, a fundamental open question is how to best characterise the emergence of various collective modes of interdependency among several degrees of freedom and, more broadly, which are the laws describing the informational architecture of complex systems. 
While physical systems and processes can be effectively described (either from first principles or as a modelling strategy) via pairwise interactions, many-body interactions~\cite{carleo01,bianconi01,bianconi02} is gathering the attention of several communities --- even beyond physics, including neuroscience, economics, and many others~\cite{perc01,rosas02,sedina01,zhang01,stramaglia01}. 
Crucially, it has been recently noticed that the link between the relationship of the order (e.g. pairwise vs higher-order) of mechanisms and the resulting patterns can be highly non-trivial~\cite{rosas2022}. 
Therefore, a principled understanding of the informational structure of a physical system cannot be based solely on determining its effective Hamiltonian but requires additional tools to investigate the resulting phenomena that arise from it.

Information theory and statistical mechanics provide an ideal toolbox to prove the informational relationships within complex systems~\cite{jaynes2003}, benefiting from a solid mathematical formalism for describing multivariate information and high-order interactions. 
For instance, the Partial Information Decomposition (PID)~\cite{randall01,lizier01,wibral01} framework and its extension to time-series analysis~\cite{mediano2021} provide an encompassing and thorough approach to investigate the different information modes --- e.g. synergy and redundancy --- within a system, which can yield important scientific insights. 
For instance, recent investigations on brain dynamics uncovered that synergy and redundancy~\cite{luppi01,varley01} relate to the interplay between brain segregation and integration~\cite{sporns01}: namely, redundancy speaks to the robustness of I/O in sensory areas, whilst synergy dominates high-level networks and ensures information integration.

Given the relevance of synergies and redundancies, capturing their balance is a key summary marker of the informational architecture of a system. Interestingly, while the computation of the full PID of a system may be sometimes challenging, it was shown that the calculation of the overall balance between synergy and redundancy can be done in a relatively straightforward way via a quantity named \emph{O-information}~\cite{rosas01}. Despite its recent inception, the O-information has found numerous applications~\cite{stramaglia03,gatica2021,varley2023multi} and have various theoretical extensions including pointwise~\cite{scagliarini01}, spectral~\cite{faes2022}, and harmonic decompositions~\cite{medina2021}.

Motivated by the success of this measure and building on the growing interest in studying quantum systems via the lens of partial information decomposition~\cite{enk01,acin01}, 
and in quantum information~\cite{nielsen01},
here, we investigate the feasibility of extending the \oinfo~it to quantum systems. 
To reach this goal, we map the \oinfo~to its quantum counterpart and perform a series of measures on relevant Qubit states, concentrating on $4$-Qubit systems. As shown later, this choice leads us to identify notable properties of the quantum \oinfo~and the way it can be changed using suitable time evolution operators.\\\\
\textbf{Quantum O-Information.} 
Before delving into the quantum realm, let us start by reviewing the mathematical definition of the \oinfo~for classical systems. When considering a classical system composed of $n$ degrees of freedom (or variables) denoted by $X^i$, its \oinfo~is given by~\cite{rosas01}
\begin{equation}\label{eq:oinfo}
    \Omega(\bm X) \coloneqq (n-2) H(\bm{X}) + \sum_{i=1}^{n} \big[ H(X^i) - H(\bm{X}_{-i}) \big],
\end{equation}
\noindent where $\bm{X} = (X^1, X^2, \dots, X^n)$ is the corresponding state vector, $H$ denotes Shannon's entropy, and $\bm{X}_{-i}$ is the state vector composed of all variables except $X^i$.

As demonstrated in~\cite{rosas01}, the \oinfo~describes the overall characteristic of the statistics of a system. Concretely, positive values of this quantity indicate that $\bm{X}$ is a collection dominated by redundant interdependencies, while negative values indicate $\bm{X}$ is dominated by synergistic ones.
Eventually, the \oinfo~is zero if redundancies and synergies are balanced.
For $n = 3$, $\Omega$ coincides with the interaction information~\cite{mcgill01}, a classical 
multivariate information-theoretic measure~\cite{yeung01} with interesting topological properties~\cite{baudot01}. While the interaction information represents the balance of synergy-vs-redundancy only for three variables~\cite{williams01}, the \oinfo~has this capability for all system sizes.

A critical observation is that Eq.~\eqref{eq:oinfo} relies solely on a linear combination of Shannon entropies computed on various collections of variables --- while it does not require any conditional entropy terms. 
By leveraging this, one can identify a quantum counterpart to the \oinfo~by substituting the Shannon entropy $H=-\sum p\log p$ with the von Neumann entropy $S = - \mathrm{Tr}[\rho \log \rho]$, where $\rho$ denotes the density matrix of the quantum system under consideration. 
Accordingly, we define the \emph{Quantum \oinfo} --- or, more succinctly, `\qinfo'~--- of a quantum system composed of $n$ degrees of freedom as follows: 
\begin{equation}\label{eq:q_oinfo}
    \Omega_Q(\rho) = (n-2) S(\rho) + \sum_{i=1}^{n} \big[ S(\rho_i) - S(\rho_{-i}) \big],
\end{equation}
\noindent where $\rho_i$ is the density matrix for the $i$-th component of the system, obtained by tracing out all the other degrees of freedom, whereas $\rho_{-i}$ denotes the density matrix obtained by tracing out the $i$-th component from $\rho$ and corresponds to the rest of the system.\\\\
\textbf{$n$-Qubit States.} 
A first property of the \qinfo~relates the entropy of pure states. Specifically, since $S(\rho)=0$ and $S(\rho_i) = S(\rho_{-i})$ for all $i$, pure states always have zero \qinfo.
Notwithstanding, we can measure non-zero \qinfo~values in pure states through their reduced density matrix~\cite{sommers01}, i.e. by decomposing them into subsystems.

To develop further insight about the \qinfo, let us apply it to some well-known Qubit states: the Greenberger–Horne–Zeilinger (GHZ) state~\cite{facchi01}, the $W$ state, and the maximally multipartite entangled ($MMES$) states~\cite{facchi02,facchi03}, which correspond to the eigenstates of suitable local Hamiltonians~\cite{facchi04}. Each of these states was evaluated for various sizes. 
Additionally, we also evaluated states with a fixed number of Qubits: the Yeo-Chua ($YC$) state~\cite{chua01}, the 'Hyperdeterminant' ($HD$) state~\cite{enriquez01}, and the $HS$ state~\cite{higuchi01}. 
To evaluate the \qinfo~in pure states, as before described, we calculated the \qinfo~through the density matrix obtained by tracing out one arbitrary Qubit from the pure state at hand (note that the resulting Q-info does not rely on the choice of the traced-out Qubit), which allows us to gain insight into the internal structure of these states.

The results of the evaluations --- reported in Table~\ref{table:res_size} --- show that subsystems of the $GHZ$ and the $W$ states are redundancy-dominated, and the value of the \qinfo~scales with the number of Qubits. 
In contrast, the $MMES$ are synergy-dominated, with the absolute value of the \qinfo~also increasing with the number of Qubits. 
Additionally, results also show that the $HS$ is redundancy-dominated, while the $HD$ and the $YC$ states are synergy-dominated. 
Finally, it is interesting to notice that all systems studied with $3$ Qubits exhibit zero \qinfo~--- an empirical fact that we generalise in the next section.
\begin{table}[h]
\begin{tabular}{|l|l|l|}
\hline
Qubit State & Nr Qubit & \qinfo        \\ \hline
$GHZ$ & 3 & 0.0        \\ \hline
$GHZ$ & 4 & 1.0        \\ \hline
$GHZ$ & 5 & 2.0        \\ \hline
$GHZ$ & 6 & 3.0        \\ \hline
$GHZ$ & 7 & 4.0        \\ \hline
$GHZ$ & 8 & 5.0        \\ \hline
$W$ & 3 & 0.0        \\ \hline
$W$ & 4 & 0.2451        \\ \hline
$W$ & 5 & 0.4477        \\ \hline
$W$ & 6 & 0.6087        \\ \hline
$W$ & 7 & 0.7380        \\ \hline
$W$ & 8 & 0.8438        \\ \hline
$W$ & 10 & 1.0065        \\ \hline
$W$ & 12 & 1.126       \\ \hline
$MMES$ & 3 & 0.0        \\ \hline
$MMES$ & 4 & -1.0        \\ \hline
$MMES$ & 5 & -2.0        \\ \hline
$MMES$ & 6 & -2.0        \\ \hline
$YC$ & 4 & -1.0        \\ \hline
$HD$ & 4 & -0.3491        \\ \hline
$HS$ & 4 & 0.2244        \\ \hline
\end{tabular}
\caption{The \qinfo~is reported for some quantum states on varying their number of Qubits.}\label{table:res_size}
\end{table}
\begin{figure}
    \centering
    \includegraphics[width=\columnwidth]{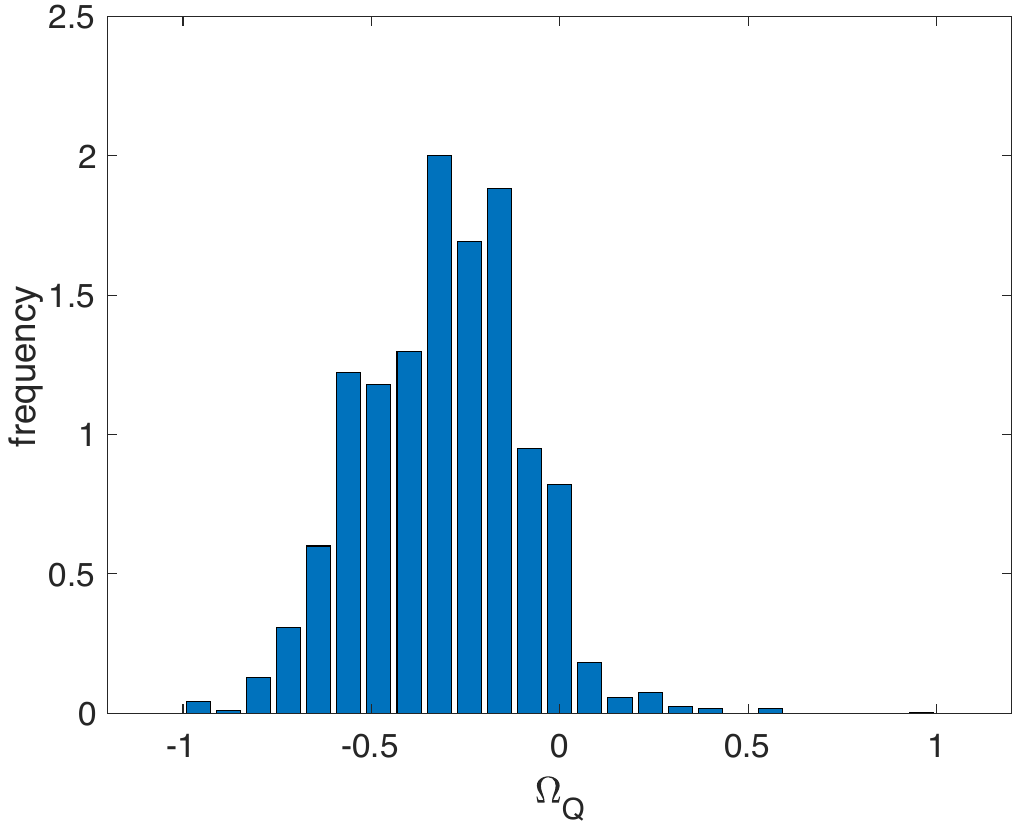}
    \caption{The distribution of the \qinfo, denoted as $\Omega_Q$, evaluated on all the $65536$ pure states that can be built by assigning $1$ or $0$ to the coefficients of the $16$ basis states. The $\Omega_Q$ ranges in $[-1, 1]$.}
    \label{fig:figure_1a}
\end{figure}
\begin{figure}
    \centering
    \includegraphics[width=\columnwidth]{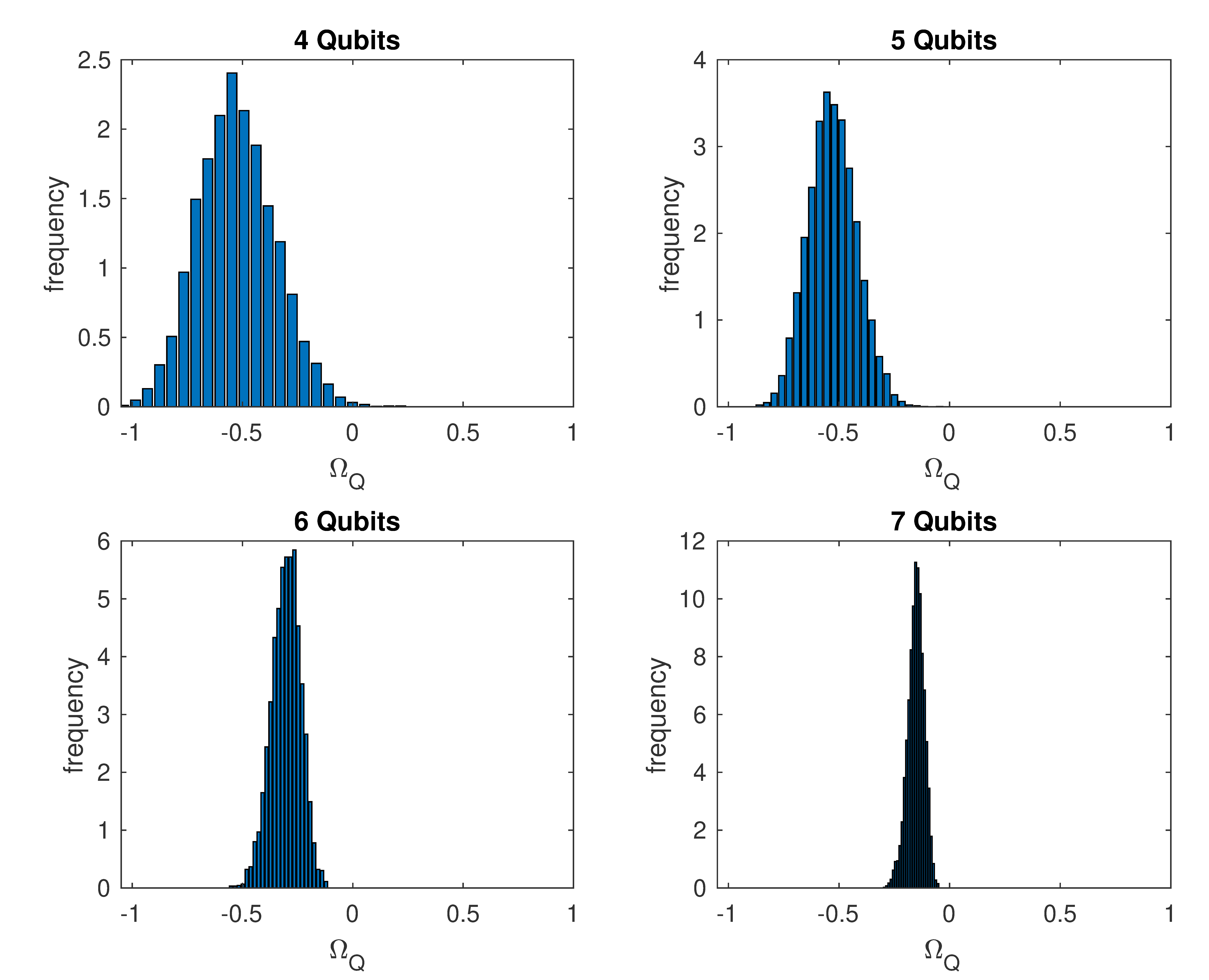}
    \caption{The distribution of the \qinfo, denoted as $\Omega_Q$, evaluated on  $8\times8$ density matrices randomly constructed as follows: $10000$ random pure states of $n$ Qubits ($n = $ $4$,$5$,$6$ and $7$) are built by firstly taking all coefficients real and imaginary parts sampled from a unit variance zero-mean Gaussian distribution. Then all coefficients are divided by the norm so as to obtain a normalized state. Eventually, the $8\times8$ density matrices are obtained by tracing out $n-3$ Qubits from the density matrix of the pure $n$-Qubit state. In the first case (n=4), the $\Omega_Q$ ranges in $[-1.1, 033]$; as $n$ increases, the measure of the $8\times8$ density matrix, induced by tracing out the Qubits, becomes more concentrated on a class of states with vanishing $\Omega_Q$, see~\cite{sommers01} for a similar concentration phenomenon in induced measure on mixed states.}
    \label{fig:figure_1b}
\end{figure}
\begin{figure*}
    \centering
    \includegraphics[scale=0.4]{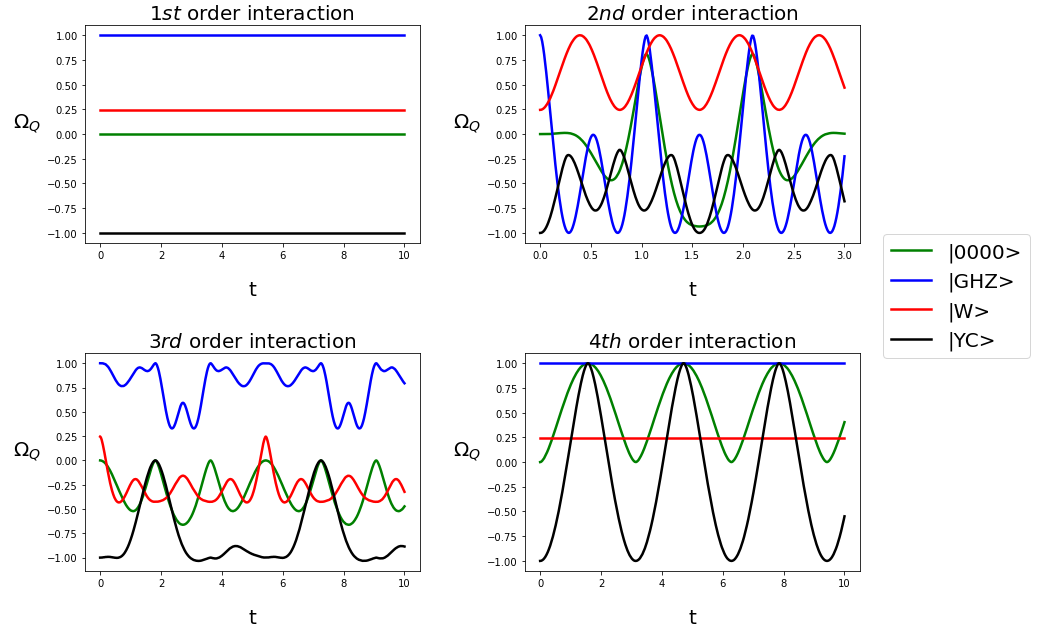}
    \caption{The time evolution of the \qinfo~depicted for the Hamiltonians defined in the main text, made of first-order terms $H_1$ (top left), second-order terms $H_2$ (top right), third-order terms $H_3$ (bottom left) and fourth-order terms $H_4$ (bottom right). The starting states are $YC$ (black line), $GHZ$ (blue line), $W$ (red line) and $\ket{0000}$ (green line).}
    \label{fig:figure_2}
\end{figure*}

To build intuition on these results, let us recall that $GHZ$ is the ground state of the ferromagnetic quantum Ising model with Hamiltonian $H = - J \sum_{\langle i,j \rangle} \sigma_i^z \tens{} \sigma_j^z$, with interaction strength $J > 0$, and $\sigma^z$ being the third Pauli matrix. 
Hence, the fact that $GHZ$ is redundancy-dominated is consistent with previous works in classical spin systems~\cite{matsuda2000physical,scagliarini01}, where the connection between synergy and frustration (i.e. when $J<0$) has been stressed. 
Other related cases of redundancy-dominated Qubit systems are states of the form $\frac{1}{\sqrt{2}}(\ket{1111}) + e^{i\alpha}\ket{0000})$ for any $\alpha$, for which $\Omega_Q=1$. 
Eventually, all the above-described properties suggest that $\Omega_Q$ departures from separableness, with a redundant character ($\Omega_Q > 0$) for entanglement typical of the $GHZ$-like states and a synergistic character ($\Omega_Q < 0$) for $MMES$-like states.\\\\
\textbf{$4$-Qubit States.} The above results suggest pure states can have a non-vanishing \qinfo~if they are composed of at least four Qubits, i.e. in these analyses, $4$-Qubit states represent the simplest non-trivial scenarios and, interestingly, connect with the concept of interaction information~\cite{mcgill01} of classical variables. Also, we remark that many open problems concern $4$-Qubit states~\cite{regula01}, e.g. their classification~\cite{ghawi01,vintskevich01,verstraete01}, and are crucial for the efficiency of quantum protocols. 
Therefore, hereinafter, we concentrate on $4$-Qubit states, whose decomposition leads to identifying four components (subsystems), say $A$, $B$, $C$, and $D$. 

Tracing out $A$, we obtain the reduced density matrix $\rho_{BCD}$.
The \qinfo~derived through the reduced density matrix $\rho_{BCD}$ reads
\begin{equation}
\label{eq:qii}
     \Omega_Q(\rho_{BCD}) = S_{BCD}-S_{BC}-S_{BD} -S_{CD}+S_{B}+S_{C}+S_{D}.
    \end{equation}
\noindent Notice that equation~\eqref{eq:qii} corresponds to the well-known topological entropy defined in~\cite{preskill01}. It is easy to prove that $\Omega_Q(\rho_{BCD})$ does not depend on the choice of $A$, as any subsystem traced out at the beginning leads to the same value. 
For instance, tracing out $B$ as a first step entails to obtain $S_{ACD}-S_{AC}-S_{AD} -S_{CD}+S_{A}+S_{C}+S_{D}$. So, from the Schmidt decomposition, we have $S_{BCD}=S(A)$, $S_{B}=S_{ACD}$, $S_{BC}=S_{AD}$, and $S_{BD}=S_{AC}$. Therefore, these equivalences allow us, by simple substitution, to prove that the \qinfo~does not depend on the choice of the traced-out variable. 
In summary, $\Omega_Q$ relies only on the quantum state and its partitioning into four components.

In light of the above observations, a first difference between the classical \oinfo~and its quantum counterpart shows up. Namely, the minimum number of degrees of freedom to have a non-vanishing \oinfo~is $n=3$ in classical systems and $n=4$ in pure quantum systems. Also, for $n=4$, there is only one way to obtain four subsystems. 
Let us remark now that a $4$-Qubit collection can have a non-vanishing $\Omega_Q$ only when its Qubits form a non-factorizable (i.e. non-separable) state.
The following bounds, that one can prove by using subadditivity~\cite{lieb01} and strong subadditivity inequalities~\cite{lieb02}, further clarify the above observation:
\begin{equation}
\label{eq:bounds}
   -2\; \min_X S_{-X}\leq  \Omega_Q \leq 2 \min_X S_{X}, 
\end{equation}
\noindent where $X$ is one of the four subsystems $A$, $B$, $C$, and $D$, while $S_{-X}$ refers to the entropy of the system measured after tracing out $X$. 
Note that if the Qubit state is factorisable (e.g. even with relation to a single Qubit), the two bounds of Eq.~\eqref{eq:bounds} vanish. 
Also, equation~\eqref{eq:bounds} entails $-6 \leq \Omega_Q \leq 2$.

Now we turn to consider all the pure states of four Qubits that can be obtained by assigning $1$ or $0$ to the coefficients of the $16$ basis states: the distribution of the $\Omega_Q$ of these states, depicted in Figure~\ref{fig:figure_1a}, shows that negative values are much more numerous. On the other hand, in Figure~\ref{fig:figure_1b} we consider systems of $4$, $5$, $6$ and $7$ Qubits and choose the expansion coefficients of the basis kets randomly and independently; then, we trace out $1$, $2$, $3$ and $4$ Qubits respectively and evaluate the $\Omega_Q$ of the resulting $8\times8$ density matrix. In all cases, the distribution of $\Omega_Q$ is peaking at negative values, i.e. we typically observe synergy. Thus, we conclude that the random generation of pure states shows that it is easier to produce synergistic Qubit configurations than redundant ones. 
Moreover, the case $n=4$ of Figure~\ref{fig:figure_1b} suggests that the bounds~\eqref{eq:bounds} are too large. Hence, as later discussed, finding better bounds represents a valuable future goal.\\\\
\textbf{Hamiltonian Time Evolution}. To capitalise on this new tool, we use the \qinfo~to address a fundamental question: can the structure of high-order interdependencies change due to the unitary time evolution of a closed quantum system? 
Since unitary operators of time evolution can alter a state vector, and the latter generates the corresponding density matrix, it is plausible that a quantum system undergoing time evolution may also experience variations in the internal organisation of information as captured by Eq.~\eqref{eq:q_oinfo}.

The time evolution operator $U$ is defined as $U = e^{-iH(t_f-t_s)}$, where $H$ is the Hamiltonian of the system of interest, and $t_f$ and $t_s$ represent the final time and the initial time of a transformation, respectively, leading between an initial and a final state.
The specific form of the Hamiltonian $H$ depends on how the system is organised. For instance, considering Qubits aligned in a one-dimensional lattice with continuous boundary conditions, a relevant choice is given by the Heisenberg model~\cite{vianna01}.
According to this model, we consider four Hamiltonians corresponding to systems with different types of interactions: first order, as given by
\begin{equation}\label{eq:heisenberg_0}
H_1 = -\sum_{i=1}^{4} \big(\sigma_{i}^{x} + \sigma_{i}^{y} + \sigma_{i}^{z}\big);
\end{equation}
\noindent second order (i.e. with pairwise interactions), as given by
\begin{equation}\label{eq:heisenberg_1}
    H_2 = -\frac{1}{2} \sum_{i=1}^{3} \big( J_x \sigma_i^x \tens{} \sigma_{i+1}^x + J_y \sigma_i^y \tens{} \sigma_{i+1}^y+ J_z \sigma_i^z \tens{} \sigma_{i+1}^z \big);
\end{equation}
\noindent and third and fourth orders (corresponding to triple and four-wise interactions), as given by
\begin{align}\label{eq:heisenberg_2}
    H_{3} =& -\frac{1}{3} \sum_{i=1}^{3} 
    \big(J_x \sigma_i^x \tens{} \sigma_{i+1}^x\tens{}\sigma_{i+2}^x 
    + J_y \sigma_i^y \tens{} \sigma_{i+1}^y\tens{}\sigma_{i+2}^y\nonumber\\
    &+ J_z \sigma_i^z \tens{} \sigma_{i+1}^z\tens{}\sigma_{i+2}^z\big),\\
    H_{4} =& -\frac{1}{4} 
    \big(J_x \sigma_1^x \tens{} \sigma_{2}^x \tens{} \sigma_{3}^x \tens{} \sigma_{4}^x 
    + J_y \sigma_1^y \tens{} \sigma_{2}^y \tens{} \sigma_{3}^y \tens{} \sigma_{4}^y \nonumber\\
    &+ J_z \sigma_1^z \tens{} \sigma_{2}^z \tens{} \sigma_{3}^z \tens{} \sigma_{4}^z\big).
\end{align}
\noindent In all the above equations, the $J$-th interaction terms can be set to $1$ without loss of generality.

We used the above Hamiltonians to investigate what types of interactions allow dynamical reconfigurations of high-order phenomena --- please notice that the relationship between the order of the mechanisms and the order of the resulting patterns can be highly non-trivial~\cite{rosas2022}. 
For this purpose, we measured the \qinfo~of $U\ket{\Psi(t)}$ at various time points, considering $U$ arising from Hamiltonians of different orders. 
Our analyses include the following entangled states $\ket{0000}$, $\ket{GHZ}$, $\ket{W}$, and $\ket{YC}$ used as initial conditions and the outcomes are shown in Figure~\ref{fig:figure_2}.

As expected, the evolution via the first-order Hamiltonian $H_1$ resulted in no variations of \qinfo, despite the non-trivial evolution induced on the Qubit state vectors (see Figure~\ref{fig:figure_2}a). 
In contrast, the time-evolution induced by the second-order Hamiltonian $H_2$ (Eq.~\eqref{eq:heisenberg_1}) leads to significant reorganisation of high-order interdependencies (see Figure~\ref{fig:figure_2}b). 
Remarkably, under these dynamics, all states alter synergy- with redundancy-dominated configurations and vice versa. Similar observations apply for the third-order Hamiltonian, i.e. $H_3$. Eventually, $H_4$ leaves unaltered the states originating from $GHZ$ and $W$ states (Figure~\ref{fig:figure_2}).
Alternations observed through these Hamiltonians show an oscillatory behaviour of \qinfo, which can be limited to a small range or even turn a synergistic configuration into a redundant one (and vice versa).
In summary, results suggest that unitary time evolution can deeply affect the internal organisation of information in entangled Qubit states.\\\\
\textbf{Conclusion.} 
Summarising, we have extended the O-information, successfully applied so far in a variety of complex 'macroscopic' systems (e.g. see~\cite{rosas03,stramaglia03,stramaglia02,scagliarini01}), to the quantum case, showing its usefulness in grasping the informational character of the internal dependencies among the components of pure quantum states. The $GHZ$ states represent the prototype of redundant dependency, while the $MMES$ states represent the synergistic prototypes. Then, after analysing in detail the $4$-Qubit case, we have found that randomly constructed pure states are typically synergistic. Eventually, we have observed that unitary time evolution performed through suitable Hamiltonians affects the internal organisation of entangled states.
In other words, we have observed that the time evolution operator, beyond increasing the quantum complexity of a state (see~\cite{jefferson01,susskind01,haferkamp01,javarone01}), can drastically change the internal information organisation of quantum systems.

To conclude, understanding the relationship between time evolution and the dynamics of \qinfo, as well as with other quantum phenomena, e.g. the entanglement monogamy~\cite{winter01}, and generalising the properties of \qinfo~to collections with $n >4$ Qubits constitute points we deem particularly relevant as future work. 
\acknowledgments
MAJ was supported by the PNRR NQST (Code: $PE23$). 
FR was supported by the Fellowship Programme of the Institute of Cultural and Creative Industries of the University of Kent, and the DIEP visitors programme at the University of Amsterdam. PF, SP and SS were supported by the Italian funding within the ``Budget MIUR - Dipartimenti di Eccellenza 2023 - 2027'' (Law 232, 11 December 2016) - Quantum Sensing and Modelling for One-Health (QuaSiModO), CUP:H97G23000100001.

\end{document}